\documentclass[12pt,a4paper]{article}      
\usepackage{a4wide}
\usepackage{parskip}
\usepackage{amsfonts}
\usepackage{amssymb}   \usepackage[mathscr]{euscript}
\usepackage{amsthm}    \usepackage{psfrag}
\usepackage{amsbsy}    \usepackage{afterpage}
\usepackage{mathrsfs}
\usepackage{amsmath}
\usepackage{bm, url}
\usepackage[numbers,sort&compress]{natbib}
\usepackage{stmaryrd}
\usepackage{graphicx}
\newtheorem{defn}{Definition} 

\def\Mu{\mathrm{M}}

\title{Emergence is coupled to scope, not level}
\author{Alex Ryan}
\date{September 2006}      

\begin{document}             

\maketitle                   

\begin{center}\section*{Abstract}\end{center}
\begin{quote}
Since its application to systems, emergence has been explained in terms of levels of observation. This approach has led to confusion, contradiction, incoherence and at times mysticism. When the idea of level is replaced by a framework of scope, resolution and state, this confusion is dissolved. We find that emergent properties are determined by the relationship between the scope of macrostate and microstate descriptions. This establishes a normative definition of emergent properties and emergence that makes sense of previous descriptive definitions of emergence. In particular, this framework sheds light on which classes of emergent properties are epistemic and which are ontological, and identifies fundamental limits to our ability to capture emergence in formal systems.
\end{quote}

\section{Introduction}
The early development of emergence in the philosophical literature was in the context of the emergence of vitality from chemistry, and the emergence of minds from biology. This promoted the importance of understanding emergence, as a potential explanation not only of the relation between the general and special sciences, but also of the evolution of life, intelligence and complexity. Yet with its dual edge, the sword of the Emergentist philosophers carved out an overly ambitious research agenda, which set as its subject domain processes that are still largely impenetrable to science. Section \ref{secHis} briefly reviews this history, a discourse that has largely obscured the fact that at heart, emergent properties are simply a difference between global and local structure.

The purpose of this paper is to advance a new definition of emergent properties and emergence. Emergence is an essential pillar of every systems approach, and yet no precise, well defined account of emergence has achieved any level of consensus among systems researchers. The current surge of interest in complex systems -- arguably the systems approach that seeks the closest integration with science -- desperately lacks a clear understanding of emergence. For systems theory to be relevant to empirical experimentation, exact, testable concepts are necessary. It is not unreasonable to expect that achieving coherence on what it means for a system to have emergent properties, and when this counts as emergence, would lead to significant advances in systems research. We suggest that the current murkiness surrounding the central concepts in complex systems represents a serious impediment to progress. Clarifying emergence is an important step towards enhancing communication within the systems community. Even more importantly, it can improve communication with other fields of inquiry, enabling -- among other possibilities -- the application of exact systems concepts in science.

The approach taken in this study departs from the tradition of philosophical fascination in the emergence of life, consciousness and the universe. Undoubtedly, the most interesting exemplars of emergence are complex, highly evolved, and probably even self-organising. However, for emergence to be a useful and unambiguous distinction, firstly it must be isolable in a more basic form. Secondly, it must be understood in terms of well defined primitives. In Section \ref{secLev}, we find that the conventional account in terms of levels and hierarchy does not meet our second criterion, so an alternative framework of scope, resolution and state is defined. In Section \ref{secProp}, an emergent property is defined, and simple examples show that novel emergent properties are coupled to scope. Section \ref{secEmergence} defines emergence as the process whereby novel emergent properties are created, and examines the relationship between emergence and predictability. Section \ref{secRethinking} uses the definition of an emergent property to outline a principled approach to determining the boundary of a system. We consider the practical limitations of the definitions in Section \ref{secLim}, and point towards some practical applications in Section \ref{secAppl}.

\section{A short history of emergence} \label{secHis}
The notion of an emergent effect was first coined in 1875 by the philosopher George Lewes\footnote{A precursor to emergence was the idea that the whole is more than the sum of its parts, which is usually attributed to Plato or Aristotle \cite{Ari04, Har02}. Lewes was also influenced by Mill's \cite{Mil43} description of heteropathic effects, whose multiple simultaneous causes were not additive.} \cite{Lew75} to describe non-additive effects of causal interactions, to be contrasted with resultants. According to Lewes,
\begin{quote}
Although each effect is the resultant of its components, we cannot always trace the steps of the process, so as to see in the product the mode of operation of each factor. In the latter case, we propose to call the effect an emergent. It arises out of the combined agencies, but in a form which does not display the agents in action ... Every resultant is either a sum or a difference of the co-operant forces; their sum, when their directions are the same -- their difference when their directions are contrary. Further, every resultant is clearly traceable in its components, because these are homogeneous and commensurable ... It is otherwise with emergents, when, instead of adding measurable motion to measurable motion, or things of one kind to other individuals of their kind, there is a cooperation of things of unlike kinds ... The emergent is unlike its components in so far as these are incommensurable, and it cannot be reduced to their sum or their difference.
\end{quote}

Several new theories of emergence appeared in the 1920s. Lewes' emergents were subtly refined by Alexander \cite{Ale20, Ale20a}, Broad \cite{Bro25} and Lloyd Morgan \cite{Llo23, Llo26, Llo33}, among others, to support a layered view of nature. The concept of emergence as a relation between simultaneous causes and their joint effect was translated to consider the upward causation of composition. Emergence now focused on properties rather than dynamical interactions, by considering the relationship between components and the whole they compose. This view on emergence was held in contrast to reductionist mechanism, the ideal that all apparently different kinds of matter are the same stuff, differing only in the number, arrangement and movement of their constituent components \cite[p45]{Bro25}. Although properties of a complex whole were still simultaneously caused by the properties of components, novel system properties were said to emerge if they could not, even in theory, be deduced from complete knowledge of the properties of the components, either taken separately or in other combinations. Emergent properties were therefore irreducible, and represented barriers to mechanistic explanations. It was this conception of emergence that became the kernel of the mid twentieth century systems movement, as summarised by Checkland \cite{Che81}:
\begin{quote}
It is the concept of organized complexity which became the subject matter of the new discipline `systems'; and the general model of organized complexity is that there exists a hierarchy of levels of organization, each more complex than the one below, a level being characterized by emergent properties which do not exist at the lower level. Indeed, more than the fact that they `do not exist' at the lower level, emergent properties are \emph{meaningless} in the language appropriate to the lower level.
\end{quote}

Although Checkland suggests that the levels of hierarchy are ordered by complexity, he is in fact defining an emergence hierarchy\footnote{An emergence hierarchy is a system view from a structural perspective made on the basis of the existence of emergent properties \cite{Bur06}.}, and there is no necessary condition on higher levels of organisation having greater complexity\footnote{Bar-Yam \cite[p5, p746]{Bar97} provides several examples of emergent simplicity. Whether higher levels of an emergence hierarchy are necessarily more complex is dependent on whether the hierarchy is nested, and if the scope and resolution of description is the same at all levels. Checkland is by no means the only author to conflate an emergence hierarchy with increasing complexity; this has been common practice since the 1920s.}. In any case, Checkland does explain the standard argument for emergence clearly. In this account, the macro language contains concepts that are meaningless at the micro level, in the way that it is meaningless to talk about flocking as a property of a single bird. It is important to emphasise that this does not claim we cannot map between sets of microstate descriptions to macrostates: it is a fundamental methodological assumption of science that such a mapping is in principle possible\footnote{The alternative position is a form of ontological pluralism, such as Cartesian substance dualism or organismic vitalism, which discourages scientific investigation. This approach declares at least one explanatory primitive (the mind or \emph{\'{e}lan vital}) which is by definition non-material and inaccessible to science.}. Instead, it is the weaker assertion that when any component of a system is viewed \emph{in isolation}, its microstate description cannot map to the associated emergent properties.

Since the systems movement adopted the conception of emergence as a relation between levels, explanations of emergence have diverged to include a remarkable number of contradictory positions. This includes a number of reductionist scientific explanations that erase the distinction between emergent properties and mechanistic explanations, relegating emergence to the merely epiphenomenal\footnote{See the survey on The Laws of Emergence in \cite[p 24]{Cor02}.}. Many other explanations tie emergence to evolution, complexity and/or self-organisation, presenting a singular unintelligible knot of concepts. Meanwhile, in contemporary philosophy a spectrum of conflicting positions, broadly either epistemological or ontological approaches to emergence, have been articulated with little headway made in either camp\footnote{Matthews \cite[p203]{Mat04} describes this impasse; also see \cite{OW05} for a review.}. As the only commonality amongst the alternative positions is their failure to gain sufficient traction to generate consensus, their variety has only reinforced the status of emergence as an enigma.

\section{Replacing level with scope and resolution} \label{secLev}
The conventional explanation of emergence presented in the previous section is unsatisfactory. The use of an emergence hierarchy to account for emergent properties is alarmingly circular, given that the levels are defined by the existence of emergent properties\footnote{For a sounder (but still not explanatory) account of emergence in terms of levels see \cite{Bun77}.}. In hierarchy theory, levels are most often considered to be epistemic, although seemingly only to avoid the burden of proof that falls on an ontological position. Many hierarchy theorists prefer to remain reality-agnostic \cite{All06}. Unsurprisingly, the inconclusive nature of levels means that explanations of emergence in terms of levels of description are unable to resolve its nature -- is emergence a natural phenomenon or an artifact of the process of observation? To bypass this impediment, we need to define emergence without invoking the concept of levels, which we argue can be accomplished using scope, resolution and state\footnote{Bar-Yam \cite{Bar04} recognises the importance of scope, scale and the microstate-macrostate relation in understanding emergence, and consequently \cite{Bar04} is closer to the following account of emergence than the sources covered in Section \ref{secHis}.}.

Scope is defined by a spatial boundary. Spatial is used in the broadest sense of the word to include conceptual and formal, as well as physical spaces, provided the system has a physical manifestation (spatial refers to the set of components, in contrast to temporal, which refers to the dynamics of those components). The scope of a system representation is the set of components within the boundary between the associated system and its environment. If an observer shifts from representing the system to representing a component, such that the component is now the system of interest, the scope of observation has narrowed. Conversely, when the scope is increased to include components that were previously part of the environment, the scope has broadened. There is also a temporal dimension to scope, which defines the set of moments of time over which the system is represented. $\mathcal{S}$ denotes scope, while $\mathcal{S}(x)$ and $\mathcal{S}(\tau)$ denote only the spatial and temporal dimensions of scope respectively.

Resolution is defined as the finest spatial distinction between two alternative system configurations. If a fine (high) and a coarse (low) resolution representation have the same scope, the fine resolution can distinguish a greater number of possibilities, $n$, and therefore each state contains more (Shannon) information, $H = - \sum_{i=1}^n p_i \log(p_i) = \log(n)$, assuming all states are equiprobable. A closely related concept is scale, which is a transformation by multiplication. The connection is that as a property is scaled up (multiplied) within a system, it can be detected at coarser resolutions. The distinction (which is rarely made) is that scale is independent of how the system is represented, whereas resolution is an attribute of the representation (scale is ontological, but resolution is epistemological). Once the resolution is set, this determines the `size' of the components that comprise the system. There is also a temporal dimension to resolution, which defines the duration of a moment in time, where longer moments represent coarser (lower) resolutions. $\mathcal{R}$ denotes resolution, while $\mathcal{R}(x)$ and $\mathcal{R}(\tau)$ denote only the spatial and temporal dimensions of resolution respectively.

The state of a system is the information that distinguishes between alternative system configurations up to some resolution at one moment in time. Macrostate $\mathrm{M}$ and microstate $\mu$ denote sets of states with two different resolutions and scopes, with the following macro-to-micro relations:
\begin{align}\label{eqnR}
\mathcal{R}_{\Mu} & \le \mathcal{R}_{\mu} \\ \label{eqnS}
\mathcal{S}_{\Mu} & \ge \mathcal{S}_{\mu} \\ \label{eqnNeq}
(\mathcal{R}_{\Mu}, \mathcal{S}_{\Mu}) & \neq (\mathcal{R}_{\mu}, \mathcal{S}_{\mu})
\end{align}
Intuitively, the macrostate has either a coarser resolution or a broader scope, or both. Let $\Mu' \in \mathscr{M}_{\Mu}$ and $\mu' \in \mathscr{M}_{\mu}$ denote the sets of $\Mu'|\mu$ and $\mu'|\Mu$ respectively satisfying Eqns. \ref{eqnR}-\ref{eqnNeq}. Also note that $\Mu$ and $\mu$ represent sets of states if $\mathcal{S}(\tau)>1$. This non-standard usage of the terms enables the representation of ensembles, and allows for emergent properties to be structured in time as well as space.

There exist other factors that influence the representation of a system by an observer. They include perspective (some information at a particular resolution is hidden eg. the state of internal organs to the naked eye) and interpretation (eg. optical illusions that have multiple valid interpretations). However, we do not need to invoke these factors to account for emergence, so for simplicity they are excluded.

\section{Emergent properties} \label{secProp}
\begin{defn}[Emergent property] \label{defnEmergent}
A property is emergent iff it is present in a macrostate and it is not present in the microstate.
\end{defn}
For our purposes, it is not necessary to impose any limitations on how the presence of emergent properties are inferred, provided the same methods are available in $\Mu$ and $\mu$. The application of a set of methods designed to infer whether an emergent property is present or not constitutes a decision procedure. If the decision procedure returns a value of 1, informally we say the property has been detected. Ideally, an emergent property should be consistently present in an ensemble, to distinguish an emergent property from a statistically unlikely transient pattern (such as a recognisable image appearing in one frame of white noise). For a macrostate $\Mu$, $P_{\mu}^{\Mu}(t) = \{p_1,p_2,...,p_n\}$ is the set of emergent properties present in $\Mu$ and not present in $\mu$ at time $t$. If $\mathcal{S}_{\Mu}(\tau)>1$, then $t$ indicates the most recent moment in $\mathcal{S}_{\Mu}(\tau)$.

It follows from the definition that emergent properties must be the result of spatially or temporally extended structures, since otherwise it would be trivial to detect their presence in the microstate. By structure, we mean there is a pattern that relates the components, which implies redundancy, and therefore the description of the components is compressible. Structure means the components are `organised' in the sense Ashby \cite{Ash62} intended: communication (in some generalised sense) occurs between components to act like a constraint in the product space of possibilities. A corollary is if the components are independent, they cannot give rise to emergent properties. Consequently a Gaussian distribution is not organised, nor is it an emergent property of IID components. The law of large numbers is a statement about the loss of structure, not the emergence of new structure. Further, superpositionality, averaging and other linear operations cannot be the source of emergent properties. This is because a linear operator evaluates equally for any arrangement of the components. Because addition is commutative, linear operations capture a common feature of a set of components independent of their organisation, so the global structure is always exactly the sum of its parts. Lewe's original insight on emergents can now be restated: nonlinearity is a necessary condition for emergent properties.

So far, we have not specified which $\Mu \in \mathscr{M}_{\Mu}$ and $\mu \in \mathscr{M}_{\mu}$ were chosen as the macrostate and microstate in Defn. \ref{defnEmergent}. Given that $\exists p_i \in P_{\Mu}(t)$, we would like to know if $p_i$ is a result of a change of resolution or scope. Both cases are now considered, by holding one of the two variables equal between the macrostate and the associated microstate.

\subsection{Class I: Weak emergent properties}\label{SecWeak}
For the first case let $\mathcal{S}_{\Mu} = \mathcal{S}_{\mu}$, which by Eqns. (\ref{eqnR}) and (\ref{eqnNeq}) implies $\mathcal{R}_{\Mu} < \mathcal{R}_{\mu}$. Hence, $H(\Mu) < H(\mu)$. Let the surjective map\footnote{That such a function exists is the fundamental methodological assumption of science referred to in Section \ref{secHis}.} $\mathcal{C}:\mathscr{M}_{\mu} \rightarrow \mathscr{M}_{\Mu}$ give the coarsegrained macrostate corresponding to the microstate $\mu$. By Defn. \ref{defnEmergent}, a decision procedure exists to detect $p_i$ in $\Mu$. But if $p_i$ can be detected in $\Mu = \mathcal{C}(\mu)$, it can also be detected in $\mu$ by applying $\mathcal{C}$, followed by the decision procedure for $\Mu$. Therefore, in this case, the presence of $p_i$ in $\Mu$ implies the presence of $p_i$ in $\mu$.

The difficulty with determining the presence of emergent properties arises from finding the map $\mathcal{C}$ in the first place: it represents `hidden structure', rather than `novel structure'. This problem reduces to a combinatorial search problem for the mapping that reveals the relationship between the structure hidden in $\mu$, and its more apparent representation in $\Mu$. The worst scenario is if $\mathcal{C}$ is non-recursive, in which case the procedure for detection outlined above is incomputable. However, it is unknown whether physical processes exist that are capable of performing non-recursive mappings\footnote{See \cite{CS99} for a review of the possibility of physical processes whose behaviour conforms to non-recursive mappings.}. Aside from incomputability, the most challenging case is an incompressible iterative function, such that $\mathcal{C}$ involves a large but finite number of transformations. For example, consider a simulation of a discrete time dynamical system, where $\mu$ is a vector of the initial conditions and updating rules, and $\Mu$ is the binary terminal state of the simulation. $\mathcal{C}$ is incompressible if the most efficient way to infer $\Mu$ given $\mu$ is by running the simulation\footnote{This corresponds to Darley's \cite{Dar94} definition: ``A true emergent phenomenon is one for which the optimal means of prediction is simulation''. Holland's \cite{Hol98} book on emergence takes a very similar approach, without committing to a precise definition.}.

Properties in this case are classified as weak emergent properties, which is consistent with Bar-Yam's \cite[p17]{Bar04} definition of ``the relationship of microscopic and macroscopic views of a system that differ only in precision''. A weak emergent property is epistemic, since once we have discovered the right mapping $\mathcal{C}$, by Defn. \ref{defnEmergent} it can no longer be considered emergent. Even in the extreme case of incomputability, it is a limitation in our ability to detect the property that creates the appearance of it being emergent. In other words, we only believe a weak emergent property is not present in $\mu$ because of practical or fundamental limitations in our ability to detect and deduce the consequences of the structures that give rise to the emergent property in $\Mu$. When practical limitations are the cause, a weak emergent property may appear to be emergent to one observer, but is not emergent to an observer with a deeper understanding of the microstate\footnote{In the literature, note that the label `weak' is rarely used by proponents of this position. However, a purely epistemic conception of emergence is usually revealed by the assignment of emergence to the \emph{relationship} between the observer and the system. A good example is Weinberg \cite[p60]{Wei01}, who states ``We can always find cases in which a property will be `emergent' to one observer and `predictable' to another''.}. Within the assumptions and definitions of this study, if resolution is the only difference between a macrostate and microstate, no property of the macrostate can be genuinely emergent from the microstate.

The class of weak emergent properties can be summarised by the following definition.
\begin{defn}[Weak Emergent Property] A property is weakly emergent iff it is present in a macrostate but it is not apparent in the microstate, and this macrostate differs from the microstate only in resolution. A weak emergent property is a limitation of the observer, not a property of the system.
\end{defn}\label{defnWeak}

\subsection{Class II: Novel emergent properties}\label{secNovel}
For the second case let $\mathcal{R}_{\Mu} = \mathcal{R}_{\mu}$, which by Eqns. (\ref{eqnS}) and (\ref{eqnNeq}) implies $\mathcal{S}_{\Mu} > \mathcal{S}_{\mu}$. In this case, $H(\Mu) > H(\mu)$. We need to identify a macrostate with the smallest possible scope that still exhibits $p_i$.
\begin{defn}[Minimal Macrostate]\label{defnMin}
A macrostate $\Mu^*$ is minimal with respect to an emergent property, if the emergent property is present in $\Mu^*$, and it is not present in any $\mu$ with the same resolution and narrower scope (ie. in any proper subset of the components of $\Mu^*$).
\end{defn}
$\Mu^*$ is not necessarily unique, since if two components are interchangeable, then there are two distinct $\Mu^*$ that can both satisfy Defn. \ref{defnMin}. Three simple examples will show that $\Mu^*$ can be well defined.

\begin{figure}[!htb]
\begin{center}
\includegraphics[scale=0.5]{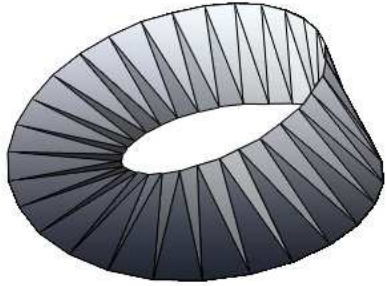}
\end{center}
\caption{A M\"{o}bius strip can be triangulated to show it has novel emergent properties.}
\label{figMoebius}
\end{figure}

Firstly, consider a M\"{o}bius strip, which is a one sided, one edged, non-orientable `surface with boundary'. We can think of the M\"{o}bius strip as being comprised of a singly twisted loop of triangles, such as the tiling depicted in Figure \ref{figMoebius}. It can be shown that any compact differentiable manifold allows a triangulation. $\mathcal{R}_{\Mu^*}$ is determined by the number of triangles used, and $\mathcal{S}_{\Mu^*}$ equals the set of triangles. If we consider any proper subset by removing at least one triangle, the resulting surface or surfaces are two sided, orientable and have more than one edge. Formally, the Euler characteristic $\chi$ is 0 for a M\"{o}bius strip, but equals the number of disjoint simplicial (triangular) complexes in $\mu$. This is equal to or greater than 1, so $\Mu^*$ is not topologically equivalent to any $\mu$. Therefore, the properties associated with the M\"{o}bius strip are emergent properties of $\Mu^*$ that do not exist for narrower scopes. Further, as $\chi$ is a topological invariant, it does not depend on the resolution of the triangulation. Hence, the emergent property is coupled to the scope of $\Mu^*$, irrespective of which particular surfaces we define as the components of $\Mu^*$.

Secondly, a perfect secret sharing scheme divides some data $D$ into $n$ pieces $D_1,...,D_n$ such that:
\begin{enumerate}
\item Knowledge of any $k$ or more $D_i$ pieces makes $D$ easily computable; and
\item Knowledge of any $k-1$ or fewer $D_i$ pieces leaves $D$ completely undetermined (in the sense that all its possible values are equally likely).
\end{enumerate}
An efficient perfect secret sharing scheme is presented in \cite{Sha79} based on polynomial interpolation, where a random polynomial is chosen of degree $k-1$ such that $q(x)=a_0+a_1 x + ... + a_{k-1} x^{k-1}$ has $a_0=D$ and $a_i \in [0,p)$ are integer and bounded by a large prime $p$. Then the keys $D_i \in \mathbb{Z}$ are generated by $D_i=q(i) \!\! \mod p, \; i=1,..,n$. Interpolation retrieves a unique value for $a_0=D$ if at least $k$ values of $q(i)$ are known, but when $k-1$ values are revealed to an opponent, all polynomials associated with possible secrets $D'$ are equally likely -- the mutual information $I(D;D')=0$. Thus when $\mathcal{R}_{\Mu} = \mathcal{R}_{\mu} = \min(|D_i-D_{i'}|) = 1$ (ie. we can distinguish between each possible $D_i \in [0,p)$) and $\mathcal{S}_{\Mu^*}=k$, then $D$ is present in $\Mu^*$ and is not present in any $\mu$. By construction, $D$ is an emergent property that is coupled to any $\Mu^*$ with a scope of $k$.

The third example presents an instance where the emergent property depends on temporal rather than spatial scope. Conceptually, there is no difference between structure extended in space or time,  except that communication can only move forwards in time, meaning that structure can only constrain possibilities that are within the `future light cone' of the first component of the structure in the temporal scope. Consider a process governed by the deterministic periodic discrete time iterative function
\begin{equation}
f(t+Pn)=f(t) \quad \forall t \in \mathbb{Z},
\end{equation}
where $P \in \mathbb{Z^+}$ is the period and $n \in \mathbb{N}$ is any multiple of $P$. The property of this sequence is translational symmetry, which is present in $\Mu^*$ when $(\mathcal{R}_{\Mu^*}(t),\mathcal{S}_{\Mu^*}(t))=(1,P+1)$, and is not present in any $\mu$ with a $\mathcal{S}_{\mu}(t) \le P$. This example is trivial, but the extension of the idea of temporal emergent properties to general discrete dynamical systems includes far more interesting structures.

This class of emergent property arises from structure that is extended over the scope of the system, which we refer to as novel emergent properties. There is a difference between local and global structure in any system that exhibits emergent novelty. This explains why emergent novelty cannot be understood or predicted by an observer whose scope is limited to only one component of a system.  When the resolution of the macrostate equals the resolution of the microstate (or the inverse mapping from the macrostate to a set of microstates is known and well defined), emergent novelty is at least in part ontological. We cannot say it is fully ontological, since some \emph{a priori} concepts are used in this framework to structure our analysis. However, the minimal macrostate has an objective property that is independent of variations in the epistemic status of an observer.

\begin{defn}[Novel Emergent Property] A property is a novel emergent property iff it is present in a macrostate but it is not present in any microstate, where the microstates differ from the macrostate only in scope.
\end{defn}\label{defnNovel}

One subclass of emergent novelty that has been discussed as a separate phenomenon in the literature is `emergent behaviour', which is a property of the system that is only exhibited in certain environments. An example of emergent behaviour is the interaction between a lock and a key\footnote{Following Bar-Yam \cite{Bar04}, we use this example for ease of comparison with what Bar-Yam calls `type 3 strong emergence' or `environmental emergence'. Additionally, a popular example of emergence in the literature is the smell of ammonia, which is said to emerge from the odourless nitrogen and hydrogen components. What is not usually explained is that the smell of ammonia is a property of the relationship between the gas and human olfactory receptors, and is therefore an example of environmental emergence. According to `shape' theories of olfaction, the interaction between ammonia molecules and the human receptor system is not dissimilar to a lock and key.}. The key is said to have an emergent behaviour, since it opens any door containing a complementary lock, and this is not present in the microstate description of the key \emph{in isolation}. This property is only present in the macrostate when the spatial scope of the system is expanded to include both the lock and the key. The emergent behaviour can then be explained as complementary spatially extended structure between two system components. From this analysis, we conclude that emergent behaviours are the result of mistakenly attributing a novel emergent property of a system to one of its components. While it is often convenient to keep the idealised system boundary fixed and talk of emergent behaviours, at the same time we should be clear that the scope of the emergent property extends between the system and certain contexts.

The classification of emergent properties has not considered the case that $\mathcal{R}_{\Mu} < \mathcal{R}_{\mu}$ and $\mathcal{S}_{\Mu} > \mathcal{S}_{\mu}$. This case is harder to analyse, since we cannot say whether $H(\Mu)$ is greater or less than $H(\mu)$. Fortunately, because we can rule out resolution as a source of emergent properties, the only important factor is scope. Therefore, we can coarsegrain $\mu$ so that $\mathcal{R}_{\Mu} = \mathcal{R}_{\mu}$, and which reduces it to Class II.

\section{Emergence}\label{secEmergence}
So far we have analysed emergent properties without saying how they arise. This is the process of emergence.
\begin{defn}[Emergence] \label{defnEmergence}
Emergence is the process whereby the assembly, breakdown or restructuring of a system results in one or more novel emergent properties.
\end{defn}
Assembly and breakdown are the dual processes of adding and removing interactions between system components that change the cardinality of the set of components in the system, while restructuring changes interactions between components without changing the cardinality. Some researchers claim that only self-assembling and self-restructuring processes can emerge. For example, Holland \cite{Hol98} argues that emergence must be the product of self-organisation, not centralised control. This is another example of attempting to tie together separate concepts that are useful only if they have distinct meanings. It usually leads to circular definitions (emergence is self-organising; self-organisation is a process that gives rise to emergence), or greedy reductions (emergence is nothing-but self-organisation).

When an emergent property is reinforced by positive feedback, its scale is increased. However, it is important to emphasise that merely scaling an emergent property is not emergence. The common use of the word does not make this distinction. For example, consider the following observation on avian bird flu. ``Despite the widespread emergence of H5N1 influenza viruses in poultry in many countries in Asia in early 2004, there were no outbreaks of H5N1 influenza in poultry or humans in Hong Kong during this time.'' \cite{LGWOPGPW04}
In this case, ``widespread emergence'' is synonymous with ``growth''. It refers to scaling of the population infected by H5N1, rather than the assembly of the initial mutation of the virus. Scaling is an important process, since if an emergent property is not reinforced it cannot perpetuate or have a significant impact (consider a mutation that does not replicate). However, the technical definition of emergence only applies to the initial process of assembly.

Having made this distinction, we can now make a useful observation on the connection between centralised control and emergence. Centralised control is characterised by a lack of autonomy in the system's components, except for the controller. The controller can assemble (or breakdown or restructure) the other components of the system, which may result in a spatiotemporally extended property. If this is performed by following some template or blueprint, we can ask whether the emergent property is present in the template. If so, then the property emerged within the controller, then was scaled up (which is not emergence) as it was realised across the system's components. If not, we can say the property emerged in the system, even though it was assembled under central control from a template. If there is no template, then the emergence also occurs in the system, not the controller. Error-free, context-insensitive asexual replication can only scale existing emergent properties, but the introduction of mutation, crossover, retroviruses or `nurture' (ie. sensitivity to information from the environment) can lead to emergence. In summary, emergence can occur through centralised control, provided the emergent property is not already present in the controller.

Emergence is defined above as a process, which means it is temporally extended. That is, emergence is not a property of a system at any point in time, it is a relationship between system properties at two different moments in time. Let $\Mu$ have emergent properties $P_{\Mu}(t)=\{p_1,p_2,...,p_n\}$ at time $t$. At some later time $t'$, the system's emergent properties are $P_{\Mu}(t')=\{p_1,p_2,...,p_{n-r},p_{n+1},...,p_{n+s} \}, \text{ with } r,s \in \mathbb{N}$. If $\max \{r, s\} > 0$, then at least one new emergent property is present, or a previous emergent property no longer exists in $\Mu$. In either case, between $t$ and $t'$, $\Mu$ exhibits emergence\footnote{Note that every $p_i$ has a logical complement, `absence of $p_i$'. Therefore the disappearance of $p_i$ is logically equivalent to the appearance of its complement, and vice versa. This is why it is not possible to separate the roles of assembly and breakdown in emergence.}.

As a thought experiment, we can set $\mathcal{S}_{\mu}$ to be unbounded in space and include all prior moments in time. In this case, $\Mu$ with the same resolution and spatial scope contains just one new moment in time. Now, if $\Mu$ contains a novel emergent property $\hat{p}_i$, then $\hat{p}_i$ has never existed before. For $\Mu^*$ such that $\hat{p}_i \in P_{\Mu^*}$, $\mathcal{S}_{\Mu^*}$ contains a corresponding structure whose specific configuration has likewise not occurred previously. We coin the term `naissance emergence' to refer to the original emergence of a novel emergent property $\hat{p}_i$. Naissance emergence is the source of novelty, and is an important distinction for a discussion of the relationship between emergence and predictability.

The problem for scientists aspiring to \emph{predict} naissance emergence is that, by definition, $\hat{p}_i$ is not present until it is within temporal scope ie. until it has already occurred! Of course, a scientist may have a theory about what properties may pertain for a configuration that has not existed. But from Section \ref{secProp}, a theory that is a linear combination of properties of the components in other subsets or configurations cannot give rise to a novel emergent property. Therefore, any theory that claims to predict naissance emergence must extrapolate $\hat{p}_i$ from a nonlinear combination of previously observed properties. But if the extrapolation is nonlinear, it is not unique. Therefore, our scientist must have multiple theories for $\hat{p}_i$, all of which are possible. There is no logical way to choose between the candidate theories, so a choice of $\hat{p}_i$ can only be justified by empirical experiment. But by conducting the experiment, $\hat{p}_i$ is now within temporal scope. Consequently, $\hat{p}_i$ cannot be predicted with certainty until it has already occurred.

This implies that formal systems, including mathematical models and computer simulations, are incapable of reproducing naissance emergence. This does not mean that once naissance emergence has occurred that we cannot alter our models to include $\hat{p}_i$ and associate it with some $\Mu^*$. It just means that we cannot do it \emph{a priori}, because we require empirical access to select between the possible properties of completely new configurations. Naissance emergence is an ontological concept, since in light of the preceding discussion it cannot be epistemic.

\section{Rethinking System Boundaries}\label{secRethinking}
Interestingly, the view of emergent properties developed in this study presents an alternative way of defining the boundary of a system. It is rare that the process of system definition is treated explicitly. However, it is suggested that the following process is typical. Firstly, the system boundary is chosen to separate the system from its environment where the interactions are weakest, which sets the scope. Weak interactions are targeted because open systems will always have flows of inputs and outputs across the boundary, but if these flows are weak compared to internal interactions, they can be either ignored or aggregated, and a systems analysis should accurately capture first order features of the system.

Secondly, once the scope of the system is set, deciding the resolution is usually straightforward. If the scope is the biosphere, it is currently infeasible to model at the resolution of individual molecules. If the scope is an individual molecule, then the resolution will need to be significantly finer if we want to say anything useful about the system. Limitations on the available cognitive and/or information processing resources provides an upper bound on the practical resolution for observing a system of any given scope, and since the upper bound is Pareto dominant with respect to information, observers could be expected to be near the Pareto frontier (ignoring extremely small scopes). Thus, limitations on information processing provides an approximately linear inversely proportional relationship between scope and resolution. Just like a camera's zoom lens, varying the scope automatically adjusts the resolution, and automatic processes are subconscious and hardwired, rather than conscious, deliberate and justified.

Thirdly, now that the scope and resolution are known, the system is composed of a finite number of components. Emergent properties belong to the system if they do not occur in the absence of the system, and are not properties of the components taken separately or in other combinations.

A number of issues arise from this kind of approach to defining a system. The most obvious problem is because it is subconscious, intuitive and unstated, it is not subject to criticism or debate. The definition of the system is axiomatic, and despite its crucial role in the success or failure of all subsequent analysis, it is placed beyond question, or rather slipped in beneath questioning. In addition, the open nature of most systems of interest means specifying a unique boundary is problematic. At what point does the flow of matter, energy and information stop being part of the environment and start being part of the system? Finally, we showed in Section \ref{secNovel} that this approach leads to the idea of emergent behaviours, whereby emergent properties are assigned to the system, when the system is only one component of the structure extending between the system and its environment that gives rise to the emergent property.

Defn. 4 enables an alternative approach for identifying the boundaries of a system. Firstly, a system is defined by a set of properties $\{p_1, p_2,...,p_n\}$ that characterise and identify that system. Secondly, for each property $i$, the minimal macrostate $\Mu_i^*$ is identified, which associates that property with a particular scope, $\mathcal{S}_{\Mu_i^*}$. Thirdly, the system boundary is defined as the set union of the scope for each property, $\bigcup_{i=1}^n \mathcal{S}_{\Mu_i^*}$. Finally, the resolution must be at least as fine as the highest resolution minimal macrostate. By starting from a set of emergent properties, the process is explicit and justified; flows from the environment are included in the system boundary only when they are a necessary component of a system property; and every property must belong to a subset of the system's components. The ontological nature of novel emergent properties means the system boundaries derived from them are not arbitrary, but reflect features of the system that are independent of the observer.

It may be argued that in practice, it is never possible to list every property of a system. This is not a major limitation due to the nature of the union operator. When a subset of the set of all system properties is considered, $\{p_1,p_2,...,p_k\}$, with $k < n$, the derived system boundary represents a lower bound on the actual system boundary. This is easy to see, since  $\bigcup_{i=1}^k \mathcal{S}_{\Mu_i^*} \subseteq \bigcup_{i=1}^n \mathcal{S}_{\Mu_i^*}$. Further, as new properties are added to the subset, the derived system boundary must converge from below to the actual system boundary. Using this procedure, we can approach a representation of a system's boundary, whose only dependence on the observer is deciding on the set of properties to be associated with the system.

\section{Practical Limitations} \label{secLim}
The analysis above has helped to clarify the role resolution and scope play in emergent properties of substantial systems with a unique, well defined microstate. Mathematical examples are useful because truth is accessible within the rules of the axiomatic system. It is possible to analytically show properties of microstates and macrostates, avoiding problems such as the theory-ladenness of observation that arise when properties must be detected empirically. Unfortunately, a number of such issues limit our ability to decisively show the presence of emergent properties and emergence in the real world. One of the most difficult aspects of identifying emergent properties in natural examples is choosing the resolution for the macrostate, which determines what are considered to be components. If a novel emergent property is present with respect to one set of components, but not for another way of defining the components, is the property really emergent? If there exists \emph{any} way of defining the components, such that the emergent property is present in a microstate with narrower scope, then the novel emergent property does not belong to the macrostate. The property is still emergent, but we have just attributed it to the wrong scope, because of our choice of resolution. In general our decision procedure cannot check every possible resolution, so in practice applying Defn. 4 could overestimate the scope of the minimal macrostate.

Microstates and macrostates are defined in Section \ref{secLev} to cater for an ensemble perspective. Although this has not been required in the examples above, except to capture temporal structure, it will often be required for physical systems. When considering sufficiently fine microstates (either quantum or semi-classical),  observations of a system over time cannot be performed on the same microstate, but rather on the ensemble of states. In this situation, a single microstate is not physically observable and therefore is not a physically meaningful concept \cite{Bar04}. This means we need to take an ensemble perspective  \cite{Bar04}, which complicates the process of observation by making it statistical, but it is still entirely compatible with the framework developed in this study.

To quote Zadeh \cite{Zah65}, ``[m]ore often than not, the classes of objects encountered in the real physical world do not have precisely defined criteria of membership''. If a property is either statistical or fuzzy, there will not be a discontinuous boundary between emergent and non-emergent. For example, a self-avoiding random walk has the property of almost completely unstructured movement when the scope of the microstate is one move, and the property of considerably structured, statistically self-similar movement when the scope of the macrostate is a large number of moves. The emergent property of statistical self-similarity is satisfied with greater confidence as the temporal scope of the macrostate broadens. In this case, the scope associated with the novel emergent property will be somewhat arbitrary. However, we can at least specify bounds on the associated scope, such that below the lower bound the novel emergent property does not exist, while it exists with an arbitrary degree of confidence above the upper bound on scope.

There exist many systems that are not studied according to the distribution of their physical substance, including formal (mathematical) systems and social systems. In these systems, a convenient property of physical systems is absent. In physical systems, entropy is well defined by the quantum difference given by Planck's constant $h$ \cite[p13]{Bar97}. This means there exists only a finite number of distinct possibilities. Even though formal and social systems must both ultimately have physical instantiations, they do not have obvious bounds (analogous to Planck's constant) on possibilities. For instance, although the number of distinct thoughts a human mind will have in its lifetime is finite, we apparently cannot specify in advance any finite set containing every possible thought, nor determine the finest possible distinction between two thoughts the mind is capable of making. In mathematical systems, the bounds on scope and resolution are even less obvious (real numbers in general contain infinite information), which is why mathematics is sometimes described as the study of all possible worlds. If the distinct possibilities are not bounded, then resolution may not be a meaningful concept, and the microstate may contain infinite information. In order to apply this framework to non-physical systems, spaces must first be approximated with a finite product space of possibilities, so that the microstate and resolution are both well defined.

\section{Practical Applications} \label{secAppl}
The preceding mathematical examples are useful because of their simplicity and precision. This section will not be precise or conclusive, but rather suggestive of real world examples of emergence. The aim is to provide a few hooks for some of the many disciplines that investigate emergent phenomena.

Simple machines such as pulleys and levers provide a mechanical advantage by decreasing the amount of force required to do a given amount of work. In isolation, simple machines increase the scale of effect of the energy source, which by Section \ref{secEmergence} is not emergence, or transform energy between forms. However, when a collection of simple machines are assembled to form a compound machine, it is possible that the compound machine has an emergent property that no proper subset of the parts and the energy source exhibit. Usually, these emergent properties are thought of as either the function of the compound machine, or unintended consequences. A particularly vivid example of emergent functionality is given by Rube Goldberg machines, named after the American cartoonist who drew improbable machines for performing simple tasks. The board game Mousetrap, and the computer game The Incredible Machine, are both instantiations of Rube Goldberg machines where no proper subset of the system components can achieve the functionality of the complete system. Almost all engineered systems -- clocks, radios, computers, and aeroplanes -- are designed for specific, predictable emergent properties. However, note that to provide robust functionality, most engineered systems contain redundancy, which means the system contains more components than the minimal macrostate. Also note that if the emergent function is a behaviour, then by Section \ref{secNovel} the emergent property is formally a property of the larger system in which the engineered system is used.

In chemistry, a catalyst decreases the activation energy of a chemical reaction. An autocatalytic set is defined as a reaction system $(M,R)$ of molecules $M$ and reactions $R$, such that all the catalysts for all its reactions $R$ are in $M$. If no proper subset of $(M,R)$ is an autocatalytic set, then the reaction system has the emergent property of catalytic closure. Some researchers, such as Kauffman \cite[p329]{Kau93}, have linked certain forms of catalytic closure with the ability to generate heritable variation, and consequently to evolve under the pressure of natural selection. Although the speculative claims of artificial chemistry have not yet been empirically demonstrated, autocatalysis is an obvious candidate for the emergence of novelty in chemistry.

In biology, many synergies have been studied that may be examples of emergence. The link between synergy and emergence has been made by Corning \cite{Cor02}, who defines synergy as ``the combined (cooperative) effects that are produced by two or more particles, elements, parts, or organisms -- effects that are not otherwise attainable''. Once again, a cautionary remark obtains that greater efficiency through synergy is just scaling, not emergence. One synergy that does appear to be due to emergent properties is obligate endosymbiosis, such as the relationship between the \emph{Olavius algarvensis} Oligochaete -- a gutless marine worm -- and the chemoautotrophic bacteria that lives inside it. Under certain conditions (such as the absence of an external source of reduced sulphur compounds), neither organism can survive in isolation. However, a syntrophic sulphur cycle recycles oxidised and reduced sulphur between the symbionts, which is believed to have enabled \emph{O. algarvensis} to colonise new habitats and extend their geographic distribution \cite{DMFBPKWETKGA01}.

Another biological example demonstrates how emergent properties can be either spatially or temporally extended. Cyanobacteria are a remarkably diverse group of prokaryotes. Different species that engage in the chemically incompatible processes of nitrogen fixing and photosynthesis have evolved different solutions to work around this obstacle. \emph{Anabaena} spatially separates the processes in separate heterocysts and passes the products between cells using filaments \cite{Sha88}. In contrast, \emph{Synechococcus} temporally separates the processes, performing photosynthesis during the day and nitrogen fixation at night \cite{DWG03}. Yet another species, \emph{Trichodesmium} both spatially and temporally separates the processes \cite{BLCKKBF01}. All three species have the same emergent property, which since it cannot be a property of a single component, must be distributed in space, time, or both space and time.

In economics, games such as the tragedy of the commons \cite{Har68} and prisoner's dilemma \cite{RC65} capture a mathematical representation of the difference between local and global structure. In these games, pursuit of a local maximum in the microstate (one player's payoff) prevents the players from maximising the macrostate (the total payoff to all players). The structure of the games and greedy rational behaviour combine to ensure a sub-optimal Nash equilibrium will prevail, even when a solution exists where all players could have received greater payoffs. This kind of outcome results because the scope of consideration of greedy rational players is too narrow. If the negative externalities of the actions of individual players are incorporated into the payoff (which broadens the scope of what each player pays attention to), then the structure of the game changes and the equilibrium will no longer be dominated by other combinations of strategies. The game is now factored, meaning any increase in the payoff to one player does not decrease the total payoff to all players \cite{WT99}. The point at which incorporating the cost of negative externalities into the payoff matrix results in a factored game is the scope associated with this emergent property. The applications of game theory are of course much wider than just economics. There are many games with a similar difference between local and global structure, such as Braess' Paradox \cite{BNW05}, where adding extra capacity to a network can reduce global performance, and Parrondo's paradox \cite{HA99}, where playing two losing games can be a winning strategy overall. In Parrondo's paradox, scope has a different meaning (playing each game in isolation or in conjunction), but the emergent property is still coupled to scope.

\section{Summary}\label{secSum}
Due to its central position in systems approaches, a redefinition of emergence has significant implications for systems research. The definitions in this study do not directly contradict the common view that emergent properties at one level are meaningless at the level below. However, they do provide a deeper understanding that represents a substantial refinement of the common conception. A simple explanation of emergent properties is given in terms of scope, which forms the basis of a normative definition of emergent properties and subsequently for emergence. That is, rather than just describing what emergent properties are like, our definition prescribes the conditions whereby a property should be formally considered to be emergent.

Defn. \ref{defnEmergent} is so general it is trivial. Given almost any macrostate, we can choose a microstate with a sufficiently small scope that the macrostate has an emergent property relative to that microstate. If almost every macrostate has emergent properties, the definition is meaningless. This is why the idea of the minimal macrostate $\Mu^*$ is crucial. $\Mu^*$ allows emergent properties to be coupled to a specific scope. Defn. 4 is more specific, because it only counts emergent properties when they are not a property of \emph{any} microstate with smaller scope. Consequently, only a small fraction of macrostates have novel emergent properties.

A number of phenomena have previously been lumped under the banner of emergence. In this study, we found that the concept is currently too broad. Weak emergent properties must be excluded from emergence: the resolution of observation, or the language of description has no bearing on whether a property is emergent. Emergent behaviour, or environmental emergence, must be reassessed as a novel emergent property of a system with larger scope. A clear distinction was made between emergent properties and emergence, which shows that simply scaling an emergent property cannot be considered emergence.

Lewes thought of emergents as the converse of resultants, while Broad recast emergentism in opposition to mechanism. In this study, an emergent property is the converse of a local property. This is consistent with Lewes, insofar as resultants represent linear combinations of existing localised components. Nonlinearity is a necessary but not sufficient condition for an emergent property. In contrast, if a property is non-local, it is spatially or temporally extended, and necessarily emergent, so this opposition is more revealing. The relationship with Broad's antonym is less clear, since it depends on how strictly mechanism is interpreted. Under the purest interpretation of mechanism, local properties should tell us everything there is to know about a system, and there is no potential for naissance emergence. However, pure mechanism is not really a serious metaphysical position, and contrasting it with emergentism does little to reveal what emergent properties are. The view of emergence in this study seems to be most compatible with non-reductive physicalism, although it is not our intent to advocate a particular ontology. In summary, Defn. \ref{defnEmergent} not only enables us to say what emergent properties are, it also allows us to better say what they are not.

As long as there exist possible configurations of our universe that have not yet occurred, we must be realistic about our ability to predict future dynamics on the basis of formal systems, when the dynamics may be influenced by emergence. This insight is the most profound implication of emergence, yet also the most difficult aspect of emergence to demonstrate constructively, by virtue of its absence in formal systems.

The alternative to level and hierarchy -- scope, resolution and state -- offers a generic framework for analysing systems. Because the primitives are well-defined for any physical system, they should have broad applicability in systems research. An obvious direction for further research is to supplement these primitives with other concepts to provide a more powerful formalism.

Given that science has always valued depth of knowledge over breadth, it is not surprising that a scientific understanding of emergence has not been forthcoming, when emergent properties are precisely the properties that cannot be understood with additional depth. The existence of emergent properties provides legitimation for broader systems approaches that complement specialised scientific disciplines. However, this insight is not articulated in discussions of emergence based on levels, because in general a level is an indeterminate mixture between relationships of scope and resolution. By revealing the coupling between emergence and scope, it is hoped that the dialogue on emergence can achieve coherence.

\section{Acknowledgements}
This research was conducted under the DSTO complex adaptive systems task, in conjunction with the CSIRO emergence interaction task \cite{Bos05}. The author gratefully acknowledges the comments of Jon Opie, Martin Burke, Fabio Boschetti and David Green.

\bibliographystyle{plain}
\bibliography{../ryanThesis}

\begin{thebibliography}{10}

\bibitem{Ale20}
S.~Alexander.
\newblock {\em Space, Time and Deity Volume 1}.
\newblock Macmillan, London, UK, 1920.

\bibitem{Ale20a}
S.~Alexander.
\newblock {\em Space, Time and Deity Volume 2}.
\newblock Macmillan, London, UK, 1920.

\bibitem{All06}
T.~F. Allen.
\newblock A summary of the principles of hierarchy theory.
\newblock \url{http://www.isss.org/hierarchy.htm}, accessed 5/6/2006, 2006.

\bibitem{Ari04}
Aristotle.
\newblock {\em Metaphysics}.
\newblock {transl. Ross, W. D. eBooks@Adelaide, originally published 350BC},
  2004.

\bibitem{Ash62}
W.~R. Ashby.
\newblock Principles of the self-organising system.
\newblock In H.~Von~Foerster and G.~W. Zopf~Jr., editors, {\em Principles of
  Self-Organization: Transactions of the University of Ilinois Symposium},
  pages 255--278, London, UK, 1962. Pergamon Press.

\bibitem{Bar97}
Y.~Bar-Yam.
\newblock {\em {Dynamics of Complex Systems}}.
\newblock Westview Press, Boulder, Colorado, 1997.

\bibitem{Bar04}
Y.~Bar-Yam.
\newblock A mathematical theory of strong emergence using multiscale variety.
\newblock {\em Complexity}, 9(6):15--24, 2004.

\bibitem{BLCKKBF01}
I.~Berman-Frank, P.~Lundgren, Y.~Chen, H.~K\"{u}pper, Z.~Kolber, B.~Bergman,
  and P.~Falkowski.
\newblock Segregation of nitrogen fixation and oxygenic photosynthesis in the
  marine cyanobacterium trichodesmium.
\newblock {\em Science}, 294(5546):1534--1537, 2001.

\bibitem{Bos05}
F.~Boschetti.
\newblock {CSIRO emergence interaction task}.
\newblock
  \url{http://www.per.marine.csiro.au/staff/Fabio.Boschetti/CSS\_emergence.htm%
}, accessed 1/9/2006, 2005.

\bibitem{BNW05}
D.~Braess, A.~Nagurney, and T.~Wakolbinger.
\newblock On a paradox of traffic planning.
\newblock {\em Transportation Science}, 39:446--450, 2005.

\bibitem{Bro25}
C.~Broad.
\newblock {\em The Mind and its Place in Nature}.
\newblock Routlegde \& Kegan Paul, London, UK, 1925.

\bibitem{Bun77}
M.~A. Bunge.
\newblock Levels and reduction.
\newblock {\em American Journal of Physiology}, 233(3):R75--R82, 1977.

\bibitem{Bur06}
M.~Burke.
\newblock Robustness, resilience and adaptability: Implications for national
  security, safety and stability (draft).
\newblock Technical report, DSTO, 2006.

\bibitem{Che81}
P.~Checkland.
\newblock {\em Systems thinking, systems practice}.
\newblock John Wiley and Sons, Chichester UK, 1981.

\bibitem{CS99}
J.~Copeland and R.~Sylvan.
\newblock Beyond the universal turing machine.
\newblock {\em Australasian Journal of Philosophy}, 77:46--66, 1999.

\bibitem{Cor02}
P.~A. Corning.
\newblock The re-emergence of ``emergence'': A venerable concept in search of a
  theory.
\newblock {\em Complexity}, 7(6):18--30, 2002.

\bibitem{Dar94}
V.~Darley.
\newblock {Emergent Phenomena and Complexity}.
\newblock {\em Alife IV}, 1994.

\bibitem{DWG03}
J.~L. Ditty, S.~B. Williams, and S.~S. Golden.
\newblock A cyanobacterial circadian timing mechanism.
\newblock {\em Annual Review of Genetics}, 37(1):513--543, 2003.

\bibitem{DMFBPKWETKGA01}
N.~Dubilier, C.~Mulders, T.~Ferdelman, D.~de~Beer, A.~Pernthaler, M.~Klein,
  M.~Wagner, C.~Erseus, F.~Thiermann, J.~Krieger, O.~Giere, and R.~Amann.
\newblock Endosymbiotic sulphate-reducing and sulphide-oxidizing bacteria in an
  oligochaete worm.
\newblock {\em Nature}, 411(6835):298--302, 2001.

\bibitem{Har68}
G.~Hardin.
\newblock {The Tragedy of the Commons}.
\newblock {\em Science}, 162(3859):1243--1248, 1968.

\bibitem{HA99}
G.~P. Harmer and D.~Abbott.
\newblock Losing strategies can win by parrondo's paradox.
\newblock {\em Nature}, 402:864, 1999.

\bibitem{Har02}
V.~Harte.
\newblock {\em {Plato on Parts and Wholes: The Metaphysics of Structure}}.
\newblock Oxford University Press, Oxford, 2002.

\bibitem{Hol98}
J.~H. Holland.
\newblock {\em Emergence : from chaos to order}.
\newblock Helix Books. Perseus Books, Cambridge, Mass., 1998.

\bibitem{Kau93}
S.~A. Kauffman.
\newblock {\em The Origins of Order: Self-Organization and Selection in
  Evolution}.
\newblock Oxford University Press, Oxford, 1993.

\bibitem{Lew75}
G.~H. Lewes.
\newblock {\em Problems of Life and Mind Vol 2}.
\newblock Kegan Paul, Trench, Turbner, \& Co., London, 1875.

\bibitem{LGWOPGPW04}
A.~S. Lipatov, E.~A. Govorkova, R.~J. Webby, H.~Ozaki, M.~Peiris, Y.~Guan,
  L.~Poon, and R.~G. Webster.
\newblock Influenza: Emergence and control.
\newblock {\em Journal of Virology}, 78(17):8951--8959, 2004.

\bibitem{Llo23}
C.~Lloyd~Morgan.
\newblock {\em Emergent Evolution}.
\newblock Williams and Norgate, London, 1923.

\bibitem{Llo26}
C.~Lloyd~Morgan.
\newblock {\em Life, Mind and Spirit}.
\newblock Williams and Norgate, London, 1926.

\bibitem{Llo33}
C.~Lloyd~Morgan.
\newblock {\em The Emergence of Novelty}.
\newblock Henry Holt and Co., New York, 1933.

\bibitem{Mat04}
D.~Matthews.
\newblock {\em Rethinking Systems Thinking: Towards a Postmodern Understanding
  of the Nature of Systemic Inquiry}.
\newblock PhD thesis, University of Adelaide, 2004.

\bibitem{Mil43}
J.~S. Mill.
\newblock {\em System of Logic}.
\newblock Longmans, Green, Reader, and Dyer, 1843.

\bibitem{OW05}
T.~O'Conner and H.~Y. Wong.
\newblock Emergent properties.
\newblock In E.~N. Zalta, editor, {\em The Stanford Encyclopedia of
  Philosophy}. The Metaphysics Research Lab, 2005.

\bibitem{RC65}
A.~Rapoport and A.~M. Chammah.
\newblock {\em Prisoner's Dilemma: A Study in Conflict and Cooperation}.
\newblock University of Michigan Press, Ann Arbor, 1965.

\bibitem{Sha79}
A.~Shamir.
\newblock How to share a secret.
\newblock {\em Communications of the ACM}, 22(11):612--613, 1979.

\bibitem{Sha88}
J.~A. Shapiro.
\newblock Bacteria as multicellular organisms.
\newblock {\em Scientific American}, 258(6):82--89, 1988.

\bibitem{Wei01}
G.~M. Weinberg.
\newblock {\em An Introduction to General Systems Thinking}.
\newblock Dorset House Publishing, New York, USA, silver anniversary edition
  edition, 2001.

\bibitem{WT99}
D.~H. Wolpert and K.~Tumer.
\newblock An introduction to collective intelligence.
\newblock Technical report, NASA Ames Research Laboratory, 1999.

\bibitem{Zah65}
L.~A. Zadeh.
\newblock Fuzzy sets.
\newblock {\em Information and Control}, 8:338--353, 1965.

\end{thebibliography}

\end{document}